\documentclass[twocolumn,english,australian,conference]{IEEEtran}
\usepackage[T1]{fontenc}
\usepackage[latin9]{inputenc}
\usepackage[letterpaper]{geometry}
\usepackage{verbatim}
\usepackage{float}
\usepackage{amsmath}
\usepackage{amsthm}
\usepackage{amssymb}
\usepackage{graphicx}

\makeatletter
\theoremstyle{plain}
\newtheorem{thm}{\protect\theoremname}
\theoremstyle{plain}
\newtheorem{lem}[thm]{\protect\lemmaname}

\usepackage{subfigure}
\usepackage{epstopdf}
\usepackage{cite}
\usepackage{citesort}
\usepackage{balance}

\makeatother

\usepackage{babel}
\addto\captionsaustralian{\renewcommand{\lemmaname}{Lemma}}
\addto\captionsaustralian{\renewcommand{\theoremname}{Theorem}}
\addto\captionsenglish{\renewcommand{\lemmaname}{Lemma}}
\addto\captionsenglish{\renewcommand{\theoremname}{Theorem}}
\providecommand{\lemmaname}{Lemma}
\providecommand{\theoremname}{Theorem}

\begin{document}

\title{Analysis of Device-to-Device Communications in Uplink Cellular Networks
with Lognormal Fading}

\author{\noindent Junnan Yang\foreignlanguage{english}{{\normalsize{}$^{\ddagger}$,}\emph{\normalsize{}
}}Ming Ding\foreignlanguage{english}{{\normalsize{}$^{\dagger}$}\emph{\normalsize{},
}}Guoqiang Mao\foreignlanguage{english}{{\normalsize{}$^{\ddagger}{}^{\dagger}$}\emph{\normalsize{}}}\\
\foreignlanguage{english}{\textit{\small{}$^{\dagger}$Data61, CSIRO,
Australia $\left\{ Email:Ming.Ding@data61.csiro.au\right\} $}}\\
\foreignlanguage{english}{\textit{\small{}$^{\ddagger}$The University
of Technology Sydney, Australia $\left\{ Email:Junnan.Yang@student.uts.edu.au,Guoqiang.Mao@uts.edu.au\right\} $}}}
\maketitle
\begin{abstract}
In this paper, using the stochastic geometry theory, we present a
framework for analyzing the performance of device-to-device (D2D)
communications underlaid uplink (UL) cellular networks. In our analysis,
we consider a D2D mode selection criterion based on an energy threshold
for each user equipment (UE). Specifically, a UE will operate in a
cellular mode, if its received signal strength from the strongest
base station (BS) is large than a threshold $\beta$. Otherwise, it
will operate in a D2D mode. Furthermore, we consider a generalized
log-normal shadowing in our analysis. The coverage probability and
the area spectral efficiency (ASE) are derived for both the cellular
network and the D2D one. Through our theoretical and numerical analyses,
we quantify the performance gains brought by D2D communications and
provide guidelines of selecting the parameters for network operations.
\end{abstract}

\begin{comment}
Device-to-Device (D2D) communication, uplink cellular network, interference
control, channel inversion power control, stochastic geometry, Log-normal
fading
\end{comment}

\section{\label{sec:Introduction}Introduction}

\selectlanguage{english}%
\begin{comment}
Placeholder
\end{comment}

\selectlanguage{australian}%
In the last decade, there has been a sharp increase in the demand
for data traffic~\cite{cisco2014global}. To address such massive
consumer demand for data communications, especially from the user
equipments (UEs) such as smartphones and tablets, many noteworthy
technologies have been proposed~\cite{7126919}, such as small cell
networks (SCNs), cognitive radio, device-to-device (D2D) communications,
etc. In particular, D2D communications are usually defined as directly
transferring data between mobile UEs which are in proximity. Due to
the short communication distance between a D2D user pair, D2D communications
hold great promise in improving network performance such as coverage,
spectral efficiency, energy efficiency, and delay. Recently, D2D underlaid
cellular networks have been standardized by the 3rd Generation Partnership
Project (3GPP). The major challenge in the D2D enabled underlaid cellular
network is the inclusion of inter-tier and intra-tier interference
due to the aggressive frequency reuse, where cellular UEs and D2D
UEs share the same spectrum resources. In parallel with the standardization
effort, recently there has been a surge of academic studies in this
area~\cite{7147772,6928445,lee2015power,6909030}.

In more detail, by using the stochastic geometry theory, Andrews,
et.al conducted network performance analyses for the downlink (DL)~\cite{6042301}
and the uplink (UL)~\cite{6516885} of SCNs, in which UEs and/or
base stations (BSs) were assumed to be randomly deployed according
to a homogeneous Poisson distribution. In~\cite{peng2014device},
Peng developed an analytical framework for the D2D underlaid cellular
network in the DL, where the Rician fading channel model was adopted
to model the small-scale fast fading for the D2D communication links.
In~\cite{7147772}, Liu provided a unified framework to analyze the
downlink outage probability in a multi-channel environment with Rayleigh
fading, where D2D UEs were selected based the received signal strength
from the nearest BS. In~\cite{7073589}, Sun presented an analytical
framework for evaluating the network performance in terms of load-aware
coverage probability and network throughput using a dynamic TDD scheme
in which mobile users in proximity can engage in D2D communications.
In~\cite{7147834}, George proposed exclusion regions to protect
cellular receivers from excessive interference from active D2D transmitters.
In~\cite{7469370}, the authors derived approximate expressions for
the distance distribution between two D2D peers conditioned on the
core network\textquoteright s knowledge of the cellular network and
analyzed the performance of network-assisted D2D discovery in random
spatial networks.

Although the existing work provides precious insights into resource
allocation and mode selection for D2D communications, there still
exists several problems:
\begin{itemize}
\item In some studies, only a single BS with one cellular UE and one D2D
pair were considered, which did not take into account the influence
from other cells.
\item The mode selection scheme in the literature was not very practical,
which was mostly based on the distance only and considered D2D receiver
UEs as an additional tier of nodes, independent of the cellular UEs
and the D2D transmitter UEs. Such tier of D2D receiver UEs without
cellular capabilities appears from nowhere and is hard to justify.
\item D2D communications usually coexist with the UL of cellular communications
due to the relatively low inter-tier interference. Such feature has
not been well treated in the literature.
\item The pathloss model is not practical, e.g., LOS/NLOS transmissions
have not been well studied in the context of D2D, and usually the
same pathloss model was used for both the cellular and the D2D tiers.
\item Shadow fading was widely ignored in the analysis, which did not reflect
realistic networks.
\end{itemize}
\selectlanguage{english}%
\begin{comment}
Placeholder
\end{comment}

\selectlanguage{australian}%
To sum up, up to now, there is no work investigating the D2D-enabled
UL cellular network with the consideration of the lognormal shadow
fading. To fill in this gap of theoretical study, in this paper, we
consider the D2D-enhanced network and develop a tractable framework
to quantify the network performance for a D2D-enabled UL cellular
network. The main contributions of this paper are summarized as follows:
\begin{itemize}
\item We introduce a hybrid network model, in which the random and unpredictable
spatial positions of mobile users and base stations are modeled as
Possion point processes. This model captures several important characteristics
of a D2D-enabled UL cellular network including lognormal fading, transmit
power control and orthogonal scheduling of cellular users within a
cell.
\item We consider a flexible D2D mode selection which is based on the maximum
DL received power from the strongest base station. Such maximum DL
signal strength based mode selection scheme helps to mitigate the
undesirable interference from D2D transmitters.
\item We present a general and analytical framework, which considers that
the D2D UEs are distributed according to a non-homogenous PPP. With
this approach, a unified performance analysis is conducted for underlaid
D2D communications and we derive analytical results in terms of the
coverage probability and the area spectral efficiency (ASE) for both
cellular UEs and D2D UEs. Our results shed new light on the system
design of D2D communications.
\end{itemize}
\selectlanguage{english}%
\begin{comment}
Placeholder
\end{comment}

\selectlanguage{australian}%
The rest of this paper is organized as follows. In Section~\ref{sec:System-model,-Assumptions,},
we introduce the system model and assumptions used in this paper.
Section~\ref{sec:Analysis} presents our main results. We provide
numerical results and more discussion in Section~\ref{sec:SIMULATION-AND-DISCUSSION}
and conclude our work in Section~\ref{sec:Conclusion}.

\section{\label{sec:System-model,-Assumptions,}System Model}

\selectlanguage{english}%
\begin{comment}
Placeholder
\end{comment}

\selectlanguage{australian}%
In this section, we present the system model that is used in this
paper.

\subsection{The Path Loss Model}

\selectlanguage{english}%
\begin{comment}
Placeholder
\end{comment}

\selectlanguage{australian}%
We consider a D2D underlaid UL cellular network, where BSs and UEs,
including cellular UL UEs and D2D UEs, are assumed to be distributed
on an infinite two-dimensional plane $\mathbf{\mathit{\mathtt{\mathbb{R}}}^{2}}$.
We assume that the cellular BSs are spatially distributed according
to a homogeneous PPP of intensity $\lambda_{b}$ , i.e., $\varPhi_{b}=\{X_{i}\}$,
where $X_{i}$ denotes the spatial locations of the $i$th BS. Moreover,
the UEs are also distributed in the network region according to another
independent homogeneous PPP $\varPhi_{u}$ of intensity $\lambda_{u}$.

The path loss functions for the UE-to-BS links and UE-to-UE links
can be captured as following
\begin{equation}
PL_{\textrm{cellular}}^{^{\textrm{dB}}}=A_{B}^{\textrm{dB}}+\alpha_{B}10\log_{10}R+\xi_{B},
\end{equation}
and
\begin{equation}
PL_{\textrm{D2D}}^{^{\textrm{dB}}}=A_{D}^{\textrm{dB}}+\alpha_{D}10\log_{10}R+\xi_{D},
\end{equation}
where the path loss is expressed in dB unit, $A_{B}^{\textrm{dB}}$
and $A_{D}^{\textrm{dB}}$ are constants determined by the transmission
frequency, $\alpha_{B}$ and $\alpha_{D}$ are path loss exponents
for the UE-to-BS links and UE-to-UE links. Moreover, we denote by
$\mathrm{\mathcal{H}_{B}}$ and $\mathrm{\mathcal{H}_{D}}$ the lognormal
fading coefficients of a CU-to-BS link and a UE-to-UE link, and we
assume that $\mathrm{\mathcal{H}_{B}=exp\left(\kappa\xi_{db}^{B}\right)}$
and $\mathrm{\mathcal{H}_{D}=exp\left(\kappa\xi_{db}^{D}\right)}$
are lognormal fading, where $\kappa=-\mathtt{\mathrm{In10/10}}$ is
a constant, .i.e., $\xi_{db}^{B}\thicksim N\left(0,\sigma_{B}{}^{2}\right)$
and $\mathrm{\xi_{db}^{D}}\thicksim N\left(0,\sigma_{D}{}^{2}\right)$.

The received power for a typical UE from a BS $b$ can be written
as
\begin{equation}
P_{b}^{\textrm{rx}}=A_{B}P_{B}\mathrm{\mathcal{H}_{B}}\left(b\right)R^{-\alpha_{B}},
\end{equation}
where $A_{B}=10^{\frac{1}{10}A_{B}^{\textrm{dB}}}$ is a constant
determined by the transmission frequency for BS-to-UE links, $P_{B}$
is the transmission power of a BS, $\mathrm{\mathcal{H}_{B}}\left(b\right)$
is the lognormal shadowing from a BS $b$ to the typical UE.

There are two modes for UEs in the considered D2D-enabled UL cellular
network, i.e., cellular mode and D2D mode. Each UE is assigned with
a mode to operate according to the comparison of the received DL power
from its serving BS with a threshold. In more detail,

\begin{equation}
Mode=\begin{cases}
\textrm{Cellular}, & \textrm{if }P^{\ast}=\underset{b}{\max}\left\{ P_{b}^{\textrm{rx}}\right\} >\beta\\
\textrm{D2D}, & \textrm{otherwise}
\end{cases},
\end{equation}
where the string variable $Mode$ takes the value of 'Cellular' or
'D2D'. In particular, for a tagged UE, if $P^{\ast}$ is large than
a specific threshold $\beta>0$, then the UE is not appropriate to
work in the D2D mode due to its potentially large interference, and
hence it should operate in the cellular mode and directly connect
with a BS. Otherwise, it should operate in the D2D mode. The UEs that
are associated with cellular BSs are referred to as cellular UEs (CU)
and the distance from a CU to its associated BS is denoted by $R^{B}$.
From~\cite{6928445} , CUs are distributed following a non-homogenous
PPP $\varPhi_{c}$. For a D2D UE, we adopt the same assumption in~\cite{7147772}
that it randomly decides to be a D2D transmitter or D2D receiver with
equal probability at the beginning of each time slot, and a D2D receiver
UE selects the strongest D2D transmitter UE for signal reception.

Base on the above system model, we can obtain the intensity of CU
as $\lambda_{c}=q\lambda_{u}$, where $q$ denotes the probability
of $P^{\ast}>\beta$ and will be derived in closed-form expressions
in Section~\ref{sec:Analysis}. It is apparent that the D2D UEs are
distributed following another non-homogenous PPP $\varPhi_{d}$, the
intensity of which is $\lambda_{d}=\left(1-q\right)\lambda_{u}$.

\subsection{The Underlaid D2D Model}

\selectlanguage{english}%
\begin{comment}
Placeholder
\end{comment}

\selectlanguage{australian}%
We assume an underlaid D2D model. That is, each D2D transmitter reuses
the frequency with cellular UEs, which incurs inter-tier interference
from D2D to cellular. However, there is no intra-cell interference
between cellular UEs since we assume an orthogonal multiple access
technique in a BS. It follows that there is only one uplink transmitter
in each cellular BS. Here, we consider a fully loaded network with
$\lambda_{u}\gg\lambda_{b}$, so that on each time-frequency resource
block, each BS has at least one active UE to serve in its coverage
area. Note that the case of $\lambda_{u}<\lambda_{b}$ is not trivial,
which even changes the capacity scaling law~\cite{Ding2017capScaling}.
Due to the page limit, we leave the study of $\lambda_{u}<\lambda_{b}$
as our future work. Generally speaking, the active CUs can be treated
as a thinning PPP $\varPhi_{c}$ with the same intensity $\lambda_{b}$
as the cellular BSs.

Moreover, we assume a channel inversion strategy for the power control
for cellular UEs, i.e.,
\begin{equation}
P_{c_{i}}=P_{0}\mathcal{\mathrm{\left(\frac{R_{i}^{\alpha_{B}}}{\mathcal{H_{\mathrm{c_{i}}}}\mathrm{A_{B}}}\right)^{\varepsilon}}},
\end{equation}
where $P_{c_{i}}$ is the transmission power of the $i$-th cellular
link, $R_{i}$ is the distance of the $i$-th link from a CU to the
target BS, $\alpha_{B}$ denotes the pathloss exponent, $\epsilon\in(0,1]$
is the fractional path loss compensation, $P_{0}$ is the receiver
sensitivity. For BS and D2D transmitters, they use constant transmmit
powers $P_{B}$ and $P_{d}$, respectively. Besides, we denote \foreignlanguage{english}{the
additive white Gaussian noise (AWGN) power} by $\sigma^{2}$.

\subsection{The Performance Metrics}

\selectlanguage{english}%
\begin{comment}
Placeholder
\end{comment}

\selectlanguage{australian}%
According to~\cite{6042301}, the coverage probability is defined
as
\begin{equation}
P_{Mode}\left(T,\lambda_{u},\alpha_{B,D}\right)=\Pr\left[\textrm{SINR}>T\right],
\end{equation}
where $T$ is the SINR threshold, the subscript string variable $Mode$
takes the value of 'Cellular' or 'D2D', and the interference in this
paper consist of the interference from both cellular UEs and D2D transmitters.

Furthermore, the area spectral efficiency(ASE) in $\textrm{bps/Hz/k\ensuremath{m^{2}}}$
for a give $\lambda_{b,u}$ can be formulated as \vspace{0.2cm}

\noindent $A_{Mode}^{\textrm{ASE}}\left(\lambda_{Mode},\gamma_{0}\right)$
\[
=\lambda_{Mode}\int_{\gamma_{0}}^{\infty}log_{2}\left(1+x\right)f_{X}\left(\lambda_{Mode},\gamma_{0}\right)dx,
\]
where $\gamma_{0}$ is the minimum working SINR for the considered
network, and $f_{X}\left(\lambda_{Mode},\gamma_{0}\right)$ is the
PDF of the SINR observed at the typical receiver for a particular
value of $\lambda_{Mode}$.

For the whole network consisting of both cellular UEs and D2D UEs,
the sum ASE can be written as

\begin{equation}
A^{\textrm{ASE}}=A_{\textrm{Cellular}}^{\textrm{ASE}}+A_{\textrm{D2D}}^{\textrm{ASE}}.
\end{equation}

\section{\label{sec:Analysis}Main Results}

\selectlanguage{english}%
\begin{comment}
Placeholder
\end{comment}

\selectlanguage{australian}%
First of all, we introduce the Equivalence Method that will be used
throughout the paper~\cite{6566864}. Based on this method, we can
transfer the strongest association scheme to the nearest BS association
scheme. More specifically, with $i$th cellular link, if we let $\overline{R}_{i}=\mathrm{\mathcal{H}_{B}^{-1/\alpha_{B}}}R_{i}^{B}$,
where $R_{i}^{B}$ is the distance separating a typical user from
its tagged strongest base station,$\bar{R}_{i}$ is the distance separating
a typical user from its tagged nearest base station in another PPP,
then the received signal power in Eq.(3) and the transmission power
in Eq.(5) are written as
\begin{equation}
P^{\ast}=\max\left\{ P_{B}\mathbb{\mathrm{A_{B}}}\left(\overline{R^{B}}_{i}\right)^{-\alpha_{B}}\right\} ,
\end{equation}
and
\begin{equation}
P_{c_{i}}=\frac{P_{0}}{A_{B}{}^{\varepsilon}}\mathcal{\mathrm{\left(\overline{R}_{i}\right)^{\alpha\varepsilon}}}.
\end{equation}

Assume that a generic fading satisfy $\mathtt{\mathbb{E_{\mathrm{\mathrm{\mathcal{H}_{B}}}}\mathrm{\left[\left(\mathrm{\mathcal{H}_{B}}\right)^{2/\alpha_{B}}\right]=\exp(\frac{2\sigma_{B}^{2}}{\alpha_{B}})<\infty}}}$.
The system which consists of a non-homogeneous PPP with densities
$\lambda$ and in which each UE is associated with the BS providing
the strongest received signal power is equivalent to another system
which consists of another non-homogenous PPPs with densities $\lambda'\left(\cdotp\right)$
and in which each UE is associated with the BS providing the smallest
path loss. Besides, densities\foreignlanguage{english}{$\lambda'\left(\cdotp\right)$}is
given by
\begin{equation}
\lambda'\left(\varepsilon\right)=\frac{d}{d\varepsilon}\Lambda\left(\left[0,\varepsilon\right]\right),
\end{equation}
where
\begin{equation}
\Lambda\left(\left[0,\varepsilon\right]\right)=\pi\lambda\varepsilon^{2}\cdot e^{\frac{2\sigma_{B}^{2}}{\alpha_{B}^{2}}}.
\end{equation}

The transformed cellular network has the exactly same performance
for the typical receiver (BS or D2D RU) on the coverage probability
with the original network, which is proved in Appendix and validated
in this paper.

\subsection{The Probability of UE Operating in the Cellular Mode}

\selectlanguage{english}%
\begin{comment}
Placeholder
\end{comment}

\selectlanguage{australian}%
In this subsection, we present our results on the probability that
the UE operates in cellular mode and the equivalence distance distributions
in cellular mode and D2D mode respectively, particularly $q$ in Lemma~\ref{lem:When-operating-under}.
The derived results will be used in the analysis of the coverage probability
later.
\begin{lem}
\label{lem:When-operating-under}When operating under the model ,the
probability that a generic mobile UE registers to the strongest BS
and operates in cellular mode is given by
\begin{equation}
q=1-\exp\left(-\pi\lambda_{B}\left(\frac{A_{B}P_{B}}{\beta}\right){}^{2/\alpha_{B}}\cdot e^{\frac{2\sigma_{B}^{2}}{\alpha_{B}}}\right),
\end{equation}
and the probability that the UE operates in D2D mode is $\left(1-q\right)$.
\end{lem}
\begin{IEEEproof}
The probability of the RSS large than the threshold is given by
\begin{equation}
P=\Pr\left[\max\left(A_{B}P_{B}\mathrm{\mathcal{H}_{B}}R^{-\alpha_{B}}\right)>\beta\right],
\end{equation}

\noindent where we use the standard power loss propagation model with
path loss exponent $\alpha_{B}$ (for UE-BS links) and $\alpha_{D}$
(for UE-UE links). The the probability that a generic mobile UE operates
in cellular mode is
\begin{eqnarray}
q & = & \mathsf{\mathit{\mathrm{1}-\Pr\left[\max\left(A_{B}P_{B}\mathrm{\mathcal{H}_{B}}R^{-\alpha_{B}}\right)\leq\beta\right]}}\nonumber \\
 & = & 1-\exp\left(-\Lambda\left(\left[0,(\frac{\beta}{A_{B}P_{B}})^{-1/\alpha_{B}}\right]\right)\right)\nonumber \\
 & = & 1-\exp\left(-\pi\lambda_{B}\left(\frac{A_{B}P_{B}}{\beta}\right){}^{2/\alpha_{B}}\cdot e^{\frac{2\sigma^{2}}{\alpha_{B}^{2}}}\right),
\end{eqnarray}
which concludes our proof.
\end{IEEEproof}
\selectlanguage{english}%
\begin{comment}
Placeholder
\end{comment}

\selectlanguage{australian}%
Note that eq(13). explicitly account for the effects of channel fading,
path loss, transmit power,spatial distribution of BSs and the RSS
threshold $\beta$. From the result, one can see that the PPP $\phi_{u}$
can be divided into two PPPs: the PPP with intensity $q\lambda_{u}$
and the PPP with intensity $(1-q)\lambda_{u}$, which consist of cellular
UEs and D2D UEs, respectively. Same with\cite{6928445}, We assume
these two PPP are independent.

\subsection{Equivalence Distance Distributions }

\selectlanguage{english}%
\begin{comment}
Placeholder
\end{comment}

\selectlanguage{australian}%
The distance $R_{i}^{B}$ from a typical user to its associate BS(maximum
downlink receive power including lognormal fading) is an important
quantity to calculate the average power. According to the Equivalence
Theorem, $\overline{R}_{i}=\mathrm{\mathcal{H}_{B}^{-1/\alpha_{B}}R_{i}^{B}}$,
each UE is associated with the BS providing the strongest received
signal power is equivalent to another distribution in which each UE
is associated with the nearest BS. In this subsection, we derived
the pdf of $\overline{R}_{i}$, and then we derived the distribution
of the distance of D2D links. We can also derive the average transmission
power of CUs using this equivalence theorem and a simple validation
is showed in this subsection.
\begin{lem}
The probability density function(pdf) of $\overline{R}_{i}$ can be
written as
\begin{eqnarray}
f_{\overline{R_{i}}}\left(r\right) & = & \frac{2\pi\lambda_{B}r\cdotp\exp\left(-\pi\lambda_{B}r^{2}\cdot e^{\frac{2\sigma_{B}^{2}}{\alpha_{B}^{2}}}+\frac{2\sigma_{B}^{2}}{\alpha_{B}^{2}}\right)}{1-\exp\left(-\pi\lambda_{B}\left(\frac{B_{B}}{\beta}\right){}^{2/\alpha_{B}}\cdot e^{\frac{2\sigma^{2}}{\alpha_{B}^{2}}}\right)},
\end{eqnarray}
where $B_{B}=A_{B}P_{B}$ is a constant.
\end{lem}
\begin{IEEEproof}
The probability density function (PDF) of $\overline{R_{i}}$ can
be derived using the simple fact that the null probability of a 2-D
Poisson process in an area A is $exp(-\lambda A)$, and we have known
that $\overline{R}_{i}\leq(\frac{\beta}{B_{B}})^{-1/\alpha_{B}}$
, which leads to Lemma 2.
\end{IEEEproof}
\begin{figure}[H]
\begin{centering}
\includegraphics[width=8cm]{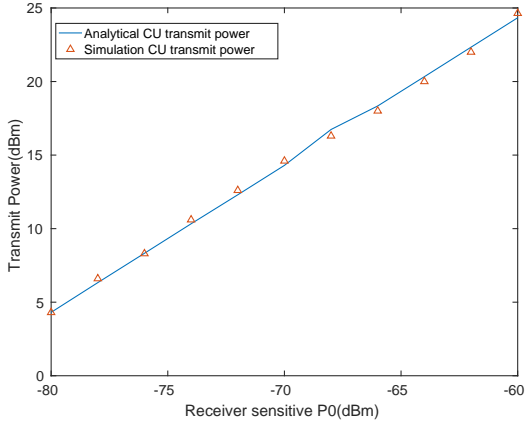}
\par\end{centering}
\caption{Transmit power of cellular UEs with $p_{0}$}
\end{figure}
As a numerical example, we plot cellular users' transmit power in
Fig.~1. The analytical result is derived from (10) and Eq.(16). It
shows that the analytical result matched well with the numerical result,
which validates our analyisis.
\begin{lem}
The typical D2D transmitter selects the equivalent nearest UE as a
potential receiver. If the potential D2D receiver is operating in
a cellular mode, D2D TU must search for another receiver. We approximate
the second neighbor as the receiver under this situation. The approximate
cumulative distribution function(CDF) of $\overline{R}_{d}$ can be
written as
\begin{align}
\Pr\left[\overline{R}_{d}<R\right]\nonumber \\
\approx & \int_{R+t}^{\infty}\left(\int_{0}^{R}f_{R_{d}}(\overline{R}_{d})d\overline{R}_{d}\right)f_{R_{d}}(r_{1})dr_{1}\nonumber \\
+ & \int_{t}^{R+t}\left(\int_{0}^{r_{1}-t}f_{R_{d}}(\overline{R}_{d})d\overline{R}_{d}\right.\\
+ & \int_{r_{1}-t}^{R}\cdot(1-Pc)\cdot f_{R_{d}}(\overline{R}_{d})d\overline{R}_{d}\nonumber \\
+ & \left.\int_{r_{1}-t}^{R}\cdot Pc\cdot f_{R_{d_{2}}}\left(\overline{R}_{d}\right)d\overline{R}_{d}\right)f_{R_{d}}(r_{1})dr_{1},\nonumber
\end{align}
where $r_{1}$is the equivalent distance from TU to the strongest
BS,$t=\left(\frac{\beta}{B_{B}}\right){}^{-1/\alpha_{B}}$. $Pc$
is the probability of a D2D receiver be a CU.
\end{lem}
\begin{IEEEproof}
If there is no different with CUs and D2D UEs, the pdf of the distance
between UEs is
\begin{equation}
f_{R_{d}}(r)=2\pi\lambda_{tu}r\cdotp\exp\left[-\pi\lambda_{tu}r^{2}\cdot e^{\frac{2\sigma_{D}^{2}}{\alpha_{D}^{2}}}+\frac{2\sigma_{D}^{2}}{\alpha_{D}^{2}}\right].
\end{equation}

Acording to\cite{1512427} , the second neighbor point is distributed
as
\begin{equation}
f_{R_{d_{2}}}(r)=2\pi^{2}\lambda_{tu}^{2}r^{3}\cdotp\exp\left[-\pi\lambda_{tu}r^{2}\cdot e^{\frac{2\sigma_{D}^{2}}{\alpha_{D}^{2}}}+\frac{4\sigma_{D}^{2}}{\alpha_{D}^{2}}\right],
\end{equation}
where $Pc$ is the probability of the potential D2D receiver operating
in cellular mode, and it can be calculated as

\begin{equation}
Pc=\arccos\left(\frac{\overline{R}_{d}+r_{1}^{2}-t^{2}}{2\overline{R}_{d}r_{1}}\right)/\pi,
\end{equation}
which concludes our proof.
\end{IEEEproof}

\subsection{Coverage Probability}

\selectlanguage{english}%
\begin{comment}
Placeholder
\end{comment}

\selectlanguage{australian}%
Consider an arbitrary BS in cellular mode or UE in D2D mode. The SINR
experenced at the receiver can be located in an arbitrary location
and can be written as
\begin{equation}
\begin{array}{c}
SINR=\frac{S_{signal}}{\underset{X_{c_{i}}\in\phi_{c}}{\sum}B_{i}^{B}\mathrm{\mathcal{H_{\mathrm{i}}^{\mathrm{B}}}}R_{B,i}^{-\alpha_{B}}+\underset{X_{d_{j}}\in\phi_{d}}{\sum}B_{j}^{D}\mathrm{\mathcal{H_{\mathrm{j}}^{\mathrm{D}}}}R_{D,j}^{-\alpha_{d}}+\eta_{c,d}}\end{array},
\end{equation}
where $B_{i}^{B}=P_{C}^{i}\cdot\textrm{\ensuremath{A_{B}}}$ and $B^{D}=P_{D}\cdot\textrm{\ensuremath{A_{D}}}$
are constant based on transmission power of the $i$th CU and the
TUs, $\mathcal{H_{\mathrm{i}}^{\mathrm{B}}}$and $\mathcal{H_{\mathrm{j}}^{\mathrm{D}}}$
are the lognorm fading in $i$th cellular uplink link and $j$th D2D
link, $R_{B,i}$and $R_{D,i}$ are the distance from the $i$th CU
and $j$th TU to the typical receiver. The Equivalence distance$\overline{R_{B}}_{i}=\mathrm{\mathcal{H}_{B,i}^{-1/\alpha_{B}}R_{B,i}}$and
$\overline{R_{D}}_{j}=\mathrm{\mathcal{H}_{D,j}^{-1/\alpha_{D}}R_{D,j}}$,
$\alpha_{B}$and $\alpha_{D}$ are path-loss exponent for cellular
links and D2D links, respectively, $\eta_{c,d}$ is the noise for
BS or receive UE.

\subsubsection{Cellular mode}

Let us consider a typical uplink, As the underlying PPP is stationnary,
without loss of generality we assume that the typical receiver is
located at the original. This analysis indicates the spatially averaged
performance of the network by Slivnyak's theorem\cite{6042301}. Henceforth,
we only need to focus on characterizing the performance of a typical
link.
\begin{lem}
The complementary cumulative distribution function(CCDF) of the SINR
at a typical BS(located in the origin)

\begin{equation}
\begin{array}{l}
\Pr\mathrm{\left[\textrm{SINR}>T\right]}\\
=\int_{0}^{\text{t}}\int_{\omega=-\infty}^{\infty}\left[\frac{e^{i\omega/T}-1}{2\pi i\omega}\right]\mathcal{F}_{SINR^{-1}}(\omega)d\omega f_{\overline{R}_{i}}(r)dr,
\end{array}
\end{equation}
where $\mathcal{F}_{SINR^{-1}}(\omega)$ denotes the conditional characteristic
function of $\frac{1}{SINR}$.
\begin{equation}
\begin{array}{l}
\mathcal{F}_{SINR^{-1}}(\omega)\\
=\exp\left\{ -2\pi\lambda_{B}e^{\frac{2\sigma^{2}}{\alpha_{B}^{2}}}\int_{t}^{\infty}\left(1-\int_{0}^{t}\exp\left(-1\times\right.\right.\right.\\
\left.\left.\left.\frac{i\omega}{(\overline{R_{B,0}})^{\alpha_{B}(\varepsilon-1)}}r^{\alpha_{B}\varepsilon}(\tau)^{-\alpha_{B}}\right)f_{\overline{R_{i}}}(r)dr\right)\tau d\tau\right\} \\
\times\exp\left\{ -\pi(1-q)\lambda_{u}e^{\frac{2\sigma^{2}}{\alpha_{B}^{2}}}\int_{t}^{\infty}\left(1-\exp\left(-1\times\right.\right.\right.\\
\left.\left.\frac{i\omega A_{B}^{\varepsilon}P_{d}}{P_{0}\cdot(\overline{R_{B,0}})^{\alpha_{B}(\varepsilon-1)}}(L)^{-\alpha_{B}}\right)LdL\right\} \\
\times\exp\left(-\frac{i\omega\eta}{\frac{P_{0}}{(\mathrm{A_{B}})^{\varepsilon-1}}\cdot(\overline{R_{B,0}})^{\alpha_{B}(\varepsilon-1)}}\right).
\end{array}
\end{equation}
\end{lem}
\begin{IEEEproof}
Conditioning on the strongest BS being at a distance $R_{B,0}$ from
the typical CU, the Equivalence distance$\overline{R_{B,0}}=\mathrm{\mathcal{H}_{B}^{-1/\alpha_{B}}R_{B,0}}$
$\left(\overline{R_{B,0}}\leq\left(\frac{\beta}{B_{B}}\right){}^{-1/\alpha_{B}}\right)$,
probability of coverage averaged over the plane is
\begin{equation}
\begin{array}{cl}
p_{c}(T,\lambda) & =\Pr[SINR>T]\\
 & =\Pr[\frac{1}{SINR}<\frac{1}{T}]\\
 & =\int_{0}^{\text{t}}\Pr[\left.\frac{1}{SINR}<\frac{1}{T}\right|\overline{R_{B,0}}]f_{\overline{R_{i}}}(r)dr
\end{array},
\end{equation}
where $i=\sqrt{-1}$ is the imaginary unit; The inner intergral is
the conditional PDF of $\frac{1}{SINR}$;$\mathcal{F}_{SINR^{-1}}(\omega)$
denotes the conditional characteristic function of $\frac{1}{SINR}$which
can be written by
\begin{eqnarray}
 &  & \mathcal{F}_{SINR^{-1}}(\omega)\nonumber \\
 & = & \mathbb{E\mathrm{_{\phi}}}\left[\left.\exp\left(-i\omega\frac{1}{SINR}\right)\right|\overline{R_{B,0}}\right]\nonumber \\
 & = & E_{\phi_{c}}\left[\exp\left(-\frac{i\omega}{\frac{P_{0}}{(\mathrm{A_{B}})^{\varepsilon-1}}\cdot(\overline{R_{B,0}})^{\alpha_{B}(\varepsilon-1)}}(I_{C})\right)\right]\nonumber \\
 & \times & \mathbb{E}{}_{\phi_{d}}\left[\exp\left(-\frac{i\omega}{\frac{P_{0}}{(\mathrm{A_{B}})^{\varepsilon-1}}\cdot(\overline{R_{B,0}})^{\alpha_{B}(\varepsilon-1)}}(I_{D})\right)\right]\nonumber \\
 & \times & \exp\left(-\frac{i\omega\eta}{\frac{P_{0}}{(\mathrm{A_{B}})^{\varepsilon-1}}\cdot(\overline{R_{B,0}})^{\alpha_{B}(\varepsilon-1)}}\right),
\end{eqnarray}
and using the definition of the Laplace transform yields, from \cite{6042301}
we have{\small{}
\begin{eqnarray}
\mathcal{L_{\mathrm{I_{c}}}\mathrm{(\mathit{s})}} & = & \mathbb{E}{}_{\phi_{c}}\left[\exp(-\mathit{sI_{c}})\right]\nonumber \\
 & = & \exp\left\{ -2\pi\lambda_{B}e^{\frac{2\sigma^{2}}{\alpha_{B}^{2}}}\int_{\overline{R_{B,0}}}^{\infty}\left(1-\int_{0}^{t}\exp\left(-1\times\right.\right.\right.\nonumber \\
 &  & \left.\left.\left.sP_{0}A_{B}^{(1-\varepsilon)}r^{\alpha_{B}\varepsilon}(\tau)^{-\alpha_{B}}\right)f_{\overline{R_{i}}}(r)dr\right)\tau d\tau\right\} ,
\end{eqnarray}
} Plugging in $s=\frac{i\omega}{\frac{P_{0}}{(\mathrm{A_{B}})^{\varepsilon-1}}\cdot(\overline{R_{B,0}})^{\alpha_{B}(\varepsilon-1)}}$
gives.{\small{}
\begin{eqnarray}
 &  & \mathbb{E}{}_{\phi_{c}}[\exp(-\frac{i\omega}{\frac{P_{0}}{(\mathrm{A_{B}})^{\varepsilon-1}}\cdot(\overline{R_{B,0}})^{\alpha_{B}(\varepsilon-1)}}(I_{C}))\nonumber \\
 & = & \exp\left\{ -2\pi\lambda_{B}e^{\frac{2\sigma^{2}}{\alpha_{B}^{2}}}\int_{t}^{\infty}\left(1-\int_{0}^{t}\exp\left(-1\times\right.\right.\right.\nonumber \\
 &  & \left.\left.\left.\frac{i\omega}{(\overline{R_{B,0}})^{\alpha_{B}(\varepsilon-1)}}r^{\alpha_{B}\varepsilon}(\tau)^{-\alpha_{B}}\right)f_{\overline{R_{i}}}(r)dr\right)\tau d\tau\right\} ,
\end{eqnarray}
}Similarly, the term $\mathbb{E}{}_{\phi_{d}}\left[\exp\left(-\frac{i\omega}{\frac{P_{0}}{(\mathrm{A_{B}})^{\varepsilon-1}}\cdot(\overline{R_{B,0}})^{\alpha_{B}(\varepsilon-1)}}(I_{D})\right)\right]$
in Eq.(25) can be written by{\small{}
\begin{eqnarray}
 &  & \mathbb{E}{}_{\phi_{d}}[\exp(-\frac{i\omega}{\frac{P_{0}}{(\mathrm{A_{B}})^{\varepsilon-1}}\cdot(\overline{R_{B,0}})^{\alpha_{B}(\varepsilon-1)}}(I_{D}))]\nonumber \\
 & = & \mathbb{E}{}_{\phi_{d}}[\underset{X_{d_{i}}\in\phi_{d}}{\prod}[\exp(-\frac{i\omega A_{B}^{\varepsilon}P_{d}}{P_{0}\cdot(\overline{R_{B,0}})^{\alpha_{B}(\varepsilon-1)}}(\overline{R_{C,i}})^{-\alpha_{B}})]]\nonumber \\
 & = & \exp\left\{ -\pi(1-q)\lambda_{u}e^{\frac{2\sigma^{2}}{\alpha_{B}^{2}}}\int_{t}^{\infty}\left(1-\right.\right.\nonumber \\
 &  & \left.\left.\exp(-\frac{i\omega A_{B}^{\varepsilon}P_{d}}{P_{0}\cdot(\overline{R_{B,0}})^{\alpha_{B}(\varepsilon-1)}}(L)^{-\alpha_{B}}\right)LdL\right\} .
\end{eqnarray}
}where $\lambda_{u}$is the intensity of Users,$R_{D,i}^{\alpha}$is
the distance from $i$th TU to typical BS.
\end{IEEEproof}

\subsubsection{Coverage Probability of D2D Mode}

\selectlanguage{english}%
\begin{comment}
Placeholder
\end{comment}

\selectlanguage{australian}%
Now let us consider a typical D2D link. As the underlying PPP is stationary,
without loss of generality, we assume that the typical receiver is
located at the original.
\begin{lem}
The CCDF of the SINR at a typical D2D UE(located in the origin)
\begin{equation}
\begin{array}{l}
\Pr[\textrm{SINR}>T]\\
=\int_{0}^{\infty}\int_{\omega=-\infty}^{\infty}\left[\frac{e^{i\omega/T}-1}{2\pi i\omega}\right]\mathcal{F}_{SINR^{-1}}(\omega)d\omega f_{\overline{R_{d}}}(r)dr,
\end{array}
\end{equation}
where $\mathcal{F}_{SINR^{-1}}(\omega)$ denotes the conditional characteristic
function of $\frac{1}{SINR}$.
\begin{equation}
\begin{array}{l}
\mathcal{F}_{SINR^{-1}}(\omega)\\
=\exp\left\{ -2\pi\lambda_{B}e^{\frac{2\sigma^{2}}{\alpha_{B}^{2}}}\int_{0}^{\infty}\left(1-\int_{0}^{t}\exp\left(-1\times\right.\right.\right.\\
\left.\left.\left.\frac{i\omega}{P_{d}(\overline{R_{d,0}})^{-\alpha_{d}}}P_{0}A_{B}^{-\varepsilon}r^{\alpha_{B}\varepsilon}(\tau)^{-\alpha_{B}}\right)f_{\overline{R_{i}}}(r)dr\right)\tau d\tau\right\} \\
\times\exp\left\{ -\pi(1-q)\lambda_{u}e^{\frac{2\sigma_{d}^{2}}{\alpha_{d}^{2}}}\int_{\overline{R_{d,0}}}^{\infty}\left(1-\exp\left(-1\times\right.\right.\right.\\
\left.\left.\frac{i\omega}{(\overline{R_{d,0}})^{-\alpha_{d}}}(L)^{-\alpha_{B}}\right)LdL\right\} \\
\times\exp\left(-\frac{i\omega\eta_{d}}{P_{d}A_{D}(\overline{R_{d,0}})^{-\alpha_{d}}}\right)
\end{array}
\end{equation}
and
\begin{equation}
f_{\overline{R_{d}}}(r)=\frac{\partial\Pr\left[\overline{R}_{d}>R\right]}{\partial\overline{R}_{d}}.
\end{equation}
\end{lem}
\begin{IEEEproof}
The proof is very similar to that for the cellular mode, and hence
we omit the proof here for brevity.
\end{IEEEproof}

\section{\label{sec:SIMULATION-AND-DISCUSSION}Simulations and Discussion}

\selectlanguage{english}%
\begin{comment}
Placeholder
\end{comment}

\selectlanguage{australian}%
In this section, we use numerical results to validate our results
on the performance of the considered D2D-enabled UL cellular network.
According to the 3GPP LTE specifications\cite{3gpp}, we set the BS
intensity to $\lambda_{B}=5\,\textrm{BSs/k\ensuremath{m^{2}}}$, which
results in an average inter-site distance of about 500$\,$m. The
UE intensity is chosen as $\lambda=300\,\textrm{UEs/k\ensuremath{m^{2}}}$~\cite{ding2015performance}.
The transmit power of each BS is $P_{B}=46\,\textrm{dBm}$, the transmit
power of D2D transmitter is $10\,\textrm{dBm}$, the path-loss exponents
are $\alpha_{c}=3.75$, $\alpha_{d}=3.75$, and the path-loss constants
are $A_{B}=10^{-3.29}$, $A_{D}=10^{-5.578}$. The threshold for selecting
cellular mode communication is set to $\beta=-65\textrm{dBm}$. The
logmormal shadowing standard deviation is $8\,\textrm{dB}$ between
UEs to BSs and $7\,\textrm{dB}$ between UEs to UEs. The noise power
is set to $-95\,\textrm{dBm}$ for a UE receiver and $-114\,\textrm{dBm}$
for a BS receiver, respectively.

\subsection{The Results on the Coverage Probability}

\begin{figure}[H]
\begin{centering}
\includegraphics[width=8cm]{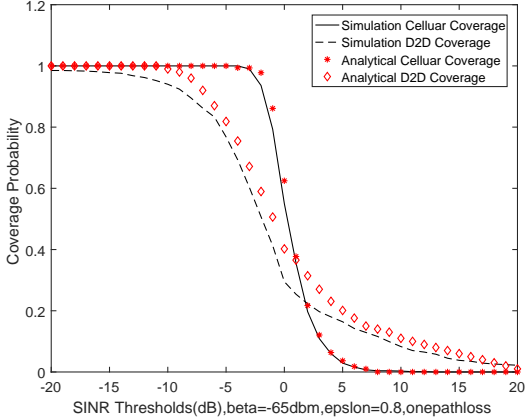}
\par\end{centering}
\caption{Coverage probabality }
\end{figure}
In Fig.2, we plot the coverage probability for both a typical cellular
UE and a typical D2D UE. From this figure, we can draw the following
observations:
\begin{itemize}
\item Our analytical results match well with the simulation results, which
validates our analysis and shows that the adopted model accurately
captures the features of D2D communications.
\item The coverage probability decreases with the increase of SINR threshold,
because a higher SINR requirement decreases the coverage probability.
\item In the D2D mode, the analytical results is shown to be larger than
the simulation resutls. This is becuase we approximate the distance
from a typical D2D TU to a typical D2D RU as that from a second nearest
D2D UE to such typical D2D RU, when the nearest D2D UE to such typical
D2D RU selects the cellular mode. However, the real distance from
a typical D2D TU to a typical D2D RU could be larger than the approximate
distance used in our analysis addressed in subsection 3.2.
\end{itemize}

\subsection{The Results on the ASE}

In Fig.3, we display the ASE results with $\gamma_{0}=0\,\textrm{dB}$.
Since $A^{\textrm{ASE}}(\lambda_{B},\lambda_{u},\gamma_{0})$ is a
function of the coverage probablity, which has been validated in Fig.2,
we only show analytical results in Fig.3.
\begin{figure}
\centering{}\includegraphics[width=8cm]{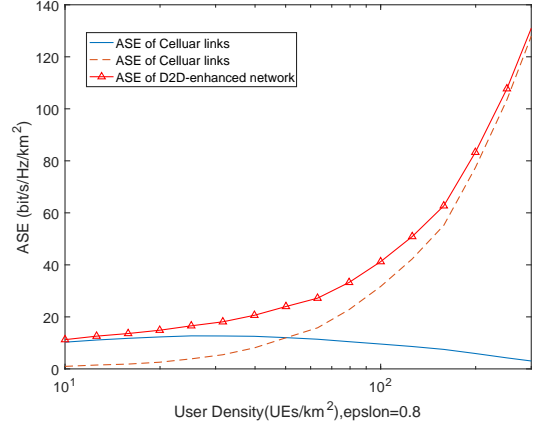}\caption{ASE with the different density of users}
\end{figure}
 From Fig.3, we can draw the following observations:
\begin{itemize}
\item The total ASE increases with the increase of the intensity of UE.
This is because the spectral reuse factor increases with the number
of UEs in the network.
\item When the intensity of UE is around $\lambda=100\,\textrm{UEs/k\ensuremath{m^{2}}}$,
the enabled-D2D links have a comparable contribution to the total
ASE as the cellular links. This is because there are around 1/3 UEs
operating in D2D mode and base on the coverage probability in D2D
tier there are around 1/3 D2D users are given a acceptable service
($\text{SINR}>0dB$), and hence they make roughly equal contributions
to the ASE performance.
\item When the network is dense enough, i.e., $\lambda_{u}\in\left[50,250\right]\textrm{UEs/k\ensuremath{m^{2}}}$,
which is the practical range of intensity for the existing 4G network
and the futrue 5G network\cite{7126919}, the total ASE performance
increases quickly, while the ASE of the cellular network stays on
top of $5\,\textrm{bps/Hz/k\ensuremath{m^{2}}}$.
\end{itemize}

\subsection{The Performance Impact of $\beta$ on the ASE}

In this subsection, we investigate the performance impact of $\beta$
on the ASE, which is shown in Fig. 4. From this figure, we can see
there is a tradeoff in the coverage probability of the cellular mode.
This means that with a proper choice of $\beta$, enabling D2D communications
not only can improve the ASE of the network, but also can improve
the coverage for cellular users.
\begin{figure}[H]
\begin{centering}
\includegraphics[width=8cm]{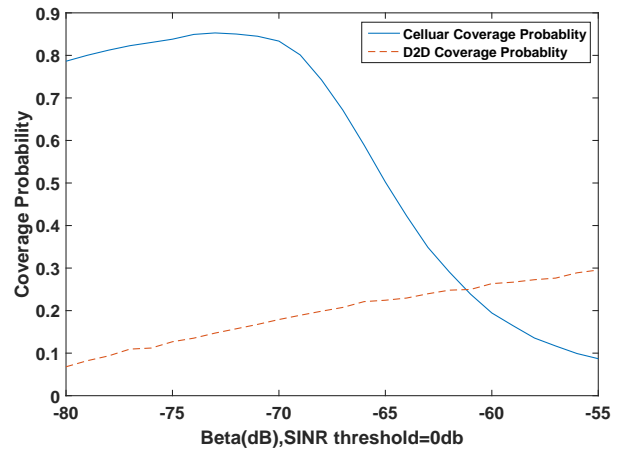}
\par\end{centering}
\caption{Coverage probability with different beta}
\end{figure}
 This is because the cell edge UEs in the conventional UL cellular
network will be offloaded to D2D modes to enjoy a better coverage
performance.

\section{\label{sec:Conclusion}Conclusion}

In this paper, we provided a stochastic geometry based theoretical
framework to analyze the performance of a D2D underlaid uplink cellular
network. In particular, we considered lognormal shadowing fading,
a practical D2D mode selection criterion based on the maximum DL received
power and the D2D power control mechanism. Our results showed that
enabling D2D communications in cellular networks can improve the total
ASE, while having a minor performance impact on the cellular network.
As future work, a more practical path loss model incorporating both
line-of-sight and non-line-of-sight transmissions will be considered,
and we will find the optimal parameters for the network that can achieve
the maximum total ASE.

\bibliographystyle{unsrt}
\bibliography{reference}

\end{document}